# Non-collinear and strongly asymmetric polar moments at back-gated SrTiO$_3$ interfaces

Fryderyk Lyzwa[1], Yurii G. Pashkevich[2], Premysl Marsik[1], Andrei Sirenko[3], Andrew Chan[4], Benjamin P.P. Mallett[4,5], Meghdad Yazdi-Rizi[1], Bing Xu[1], Luis M. Vicente-Arche[6], Diogo C. Vaz[6], Gervasi Herranz[7], Maximilien Cazayous[8], Pierre Hemme[8], Katrin Fürsich[9], Matteo Minola[9], Bernhard Keimer[9], Manuel Bibes[6], and Christian Bernhard[1]

[1]*Department of Physics and Fribourg Center for Nanomaterials, University of Fribourg, Chemin du Musée 3, CH-1700 Fribourg, Switzerland*
[2]*O. O. Galkin Donetsk Institute for Physics and Engineering NAS of Ukraine, UA-03028 Kyiv, Ukraine*
[3]*Department of Physics, New Jersey Institute of Technology, Newark, New Jersey 07102, USA*
[4]*School of Chemical Sciences and the MacDiarmid Institute for Advanced Materials and Nanotechnology, The University of Auckland, Auckland, New Zealand*
[5]*Robinson Research Institute, Victoria University of Wellington, 69 Gracefield Rd., Lower Hutt 5010, New Zealand*
[6]*Unité Mixte de Physique, CNRS, Thales, Université Paris-Saclay, 91767 Palaiseau, France*
[7]*Institut de Ciència de Materials de Barcelona (ICMAB-CSIC), Campus UAB, 08193 Bellaterra, Catalonia, Spain*
[8]*Laboratoire Matériaux et Phénomènes Quantiques (UMR 7162 CNRS), Université de Paris, 75205 Paris Cedex 13, France*
[9]*Max-Planck-Institut für Festkörperforschung, Heisenbergstrasse 1, 70569 Stuttgart, Germany*

**Abstract**

The highly mobile electrons at the interface of SrTiO$_3$ with other oxide insulators, such as LaAlO$_3$ or AlO$_x$, are of great current interest. A vertical gate voltage allows controlling a metal/superconductor-to-insulator transition, as well as electrical modulation of the spin-orbit Rashba coupling for spin-charge conversion. These findings raise important questions about the origin of the confined electrons as well as the mechanisms that govern the interfacial electric field. Here we use infrared ellipsometry and confocal Raman spectroscopy to show that an anomalous polar moment is induced at the interface that is non-collinear, highly asymmetric and hysteretic with respect to the vertical gate electric field. Our data indicate that an important role is played by the electromigration of oxygen vacancies and their clustering at the antiferrodistortive domain boundaries of SrTiO$_3$, which generates local electric and possibly also flexoelectric fields and subsequent polar moments with a large lateral component. Our results open new perspectives for the defect engineering of lateral devices with strongly enhanced and hysteretic local electric fields that can be manipulated with various other parameters, like strain, temperature, or photons.

**Introduction**

At room temperature, SrTiO$_3$ (STO) exhibits a cubic perovskite structure and a band-insulating electronic ground state with an energy gap of 3.25 eV [1], [2]. Its macroscopic properties are typical for this class of materials, except for mobile oxygen vacancies which make it an interesting ion conductor [3]. At $T^*=105$ K, STO undergoes an antiferrodistortive (AFD) transition into a tetragonal state that arises from an antiphase rotation of the TiO$_6$ octahedra around the tetragonal axis [4], [5]. If no preferred direction is imposed, e.g. by applying external pressure or electric fields, a multi-domain state develops for which the tetragonal axis is either along the x-, y-, or z-direction of the perovskite structure. The resulting AFD domain boundaries are strained and tend to be attractive for oxygen vacancies, ferroelastic and even polar [6], [7], [8], [9].

Below about 50 K, the dielectric properties of STO become highly anomalous as it approaches a ferroelectric instability that is avoided only by the quantum fluctuations of the lattice [10]. This quantum paraelectric regime is characterized by a divergence of the dielectric constant toward giant low-temperature values of up to $\varepsilon_0 \approx 20.000$ that is caused by an anomalously strong softening of the lowest infrared-active transverse phonon (TO$_1$ mode) (see e.g. [11], [12] and section 2 of the supplementary information, SI). This so-called 'soft mode' involves the off-center displacement of the titanium ions with respect to the surrounding octahedron of oxygen ions, which, in the static limit, represents the ferroelectric (FE) order parameter. The FE order can be readily induced, e.g. by isoelectronic cation substitution in Sr$_{1-x}$Ca$_x$TiO$_3$ [13], [14], via strain from a lattice mismatch with the substrate in thin films [15], by external pressure [16], by electric fields in excess of 2 kV/cm [17], and even by replacing the oxygen isotope $^{16}$O with $^{18}$O [18]. The large $\varepsilon_0$ values furthermore enhance flexoelectric effects for which strain gradients give rise to a polar moment [19].

More recently, heterostructures and devices based on STO have obtained considerable attention thanks to a great general interest in the electronic conduction along oxide-based interfaces and domain boundaries [20] as well as on the effect of oxygen intercalation on the electronic properties of metal-oxide thin films and devices, especially in the context of liquid ion gating [21]. In particular, the interface of STO with LaAlO$_3$ (LAO) or AlO$_x$ has been intensively studied because it hosts highly mobile and even superconducting electrons (below ~0.3 K [22]) that are very susceptible to electric field gating [23], [22], [24], [25], [26], [27]. The field-gating also allows tuning of the spin-orbit Rashba coupling and thus of spin-charge interconversion effects [28], [29], [30]. The origin of the confined electrons and the mechanism(s) underlying the efficient field-gating are still debated.

For LAO/STO heterostructures, the polarity of the stacking of the SrO$^0$/TiO$_2^0$/LaO$^+$/AlO$_2^-$ layers causes a diverging electric potential with increasing LAO layer thickness, which can lead to an electronic reconstruction in terms of an electron transfer from the LAO to the STO layer ('*polar catastrophe*' scenario) [24], [25]. An alternative (or additional) mechanism involves oxygen vacancies in the vicinity of the interface that have a low ionization energy, since $\varepsilon_0$ is very large [31], [32]. The latter mechanism is expected to dominate in AlO$_x$/STO, which has no obvious polar discontinuity [27], [33], [29], [28].

Here, we show with infrared ellipsometry and confocal Raman spectroscopy that for both LAO/STO (001) and AlO$_x$/STO (001) a vertical gate voltage (V) induces highly anomalous polar moments in the vicinity of the interface which are non-collinear and strongly asymmetric with respect to the nominal electric gate field. Our data indicate that the structural domain boundaries act as conduits of oxygen interstitials and thus generate large local electric fields that can greatly affect the performance of oxide devices and give rise to new functionalities.

**Infrared ellipsometry**

In infrared ellipsometry on electric-field-gated STO-based devices, an induced polar moment is manifested in characteristic changes of some of the phonon modes. This is shown in Figs. 1a and 1b for the so-called R-mode at 438 cm$^{-1}$ in AlO$_x$/STO, which exhibits a characteristic splitting at negative voltages. The R-mode becomes weakly infrared-active in the tetragonal state below $T^*=105$ K (already at 0V), where the anti-phase rotation of the neighboring TiO$_6$ octahedra leads to a backfolding from the R-point of the cubic Brillouin-zone and a mixing with the antiferroelectric displacement of the Ti ions [7], [34] as described in section 4 of the SI. In the presence of an electric-field-induced polar moment, $P$, caused by a static off-center displacement of the Ti ions, this R-mode develops a second peak (colored arrows) that is redshifted and gains spectral weight at the expense of the unshifted peak.

The right panels of Fig. 1a,b show the gate-voltage loops of the magnitude of the peak splitting, which is strongly asymmetric and vanishes at large +V. A corresponding R-mode splitting was previously observed in field-gated LAO/STO as well as in bulk SrTi$^{18}$O$_3$ where the polar moment develops in the FE state below $T^{Curie} \approx 25$ K [35]. For LAO/STO, the asymmetric gate-voltage dependence of the R-mode splitting was interpreted in terms of a built-in, vertical electric field due to the discontinuity of the polar layer stacking to which the gate field adds (is opposed) at -V (+V), such that the threshold for inducing a static polar moment is overcome (not reached). This interpretation is however challenged by our finding that this kind of R-mode splitting occurs also in AlO$_x$/STO for which no polar discontinuity is expected [27]. Moreover, based on a symmetry analysis of the infrared response of the R-mode in the presence of a polar distortion (see section 4-6 of the SI) we find that the R-mode splitting can only be seen with ellipsometry if the polar moment has a sizeable horizontal component and thus is non-collinear with respect to the vertical gate field. Our new findings therefore indicate that additional effects are at play that govern the local electric fields and the related induced polar moments in the vicinity of the AlO$_x$/STO and LAO/STO interfaces.

An anomalous origin of the induced polar moments is also suggested by the unusual training and hysteresis effects of the gate-field loops of the R-mode splitting in the right panels of Fig. 1a,b. Its onset field during the first cycle (starting from the pristine state after cooling in zero field) is strongly dependent on the sign of the variation of V. This is shown in the top-right panel of Fig. 1a where no R-mode splitting occurs as the voltage is first increased to +8 kV/cm, whereas it starts to develop around +2 kV/cm as the gate voltage is subsequently reduced. For the opposite cycle in the top-right panel of Fig. 1b, the R-mode splitting develops right away as the voltage is ramped to -V. Note that these highly asymmetric polarization and hysteresis loops are markedly different from the ones of bulk STO, for which a ferroelectric order with a symmetric hysteresis loop is induced above a threshold of about ±2 kV/cm [12].

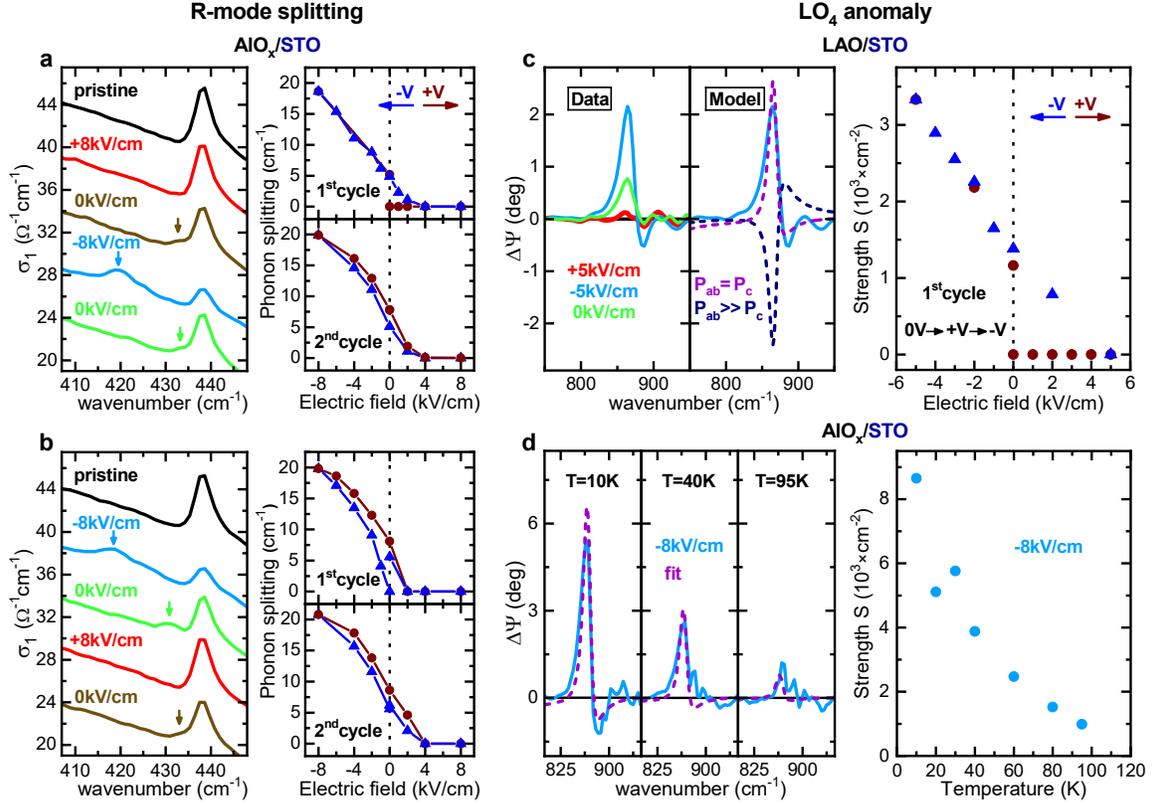

*Figure 1: Infrared ellipsometry study of phonon anomalies due to the electric-field-induced polar moment. (a) Spectra of the optical conductivity, $\sigma_1$, showing the asymmetric R-mode splitting due to $P_{ab}$ for gate-voltage cycles from 0V to +V to -V (left panel) and the corresponding hysteresis loops (right panel) for the magnitude of the R-mode splitting (and thus of $P_{ab}$) at 10 K. (b) Spectra and hysteresis loops of the R-mode splitting for the reverse voltage cycling direction, i.e. from 0V to -V to +V. (c) $LO_4$ anomaly. Left panel: The difference spectra of the ellipsometric angle $\Delta\Psi = \Psi_{exp} - \Psi_{0V,\ pristine}$ at the longitudinal optical $LO_4$ edge show the signatures of an infrared-active TO mode that is induced at -V. This new TO mode can only be modeled by assuming a polarization with a sizeable vertical component (dashed purple), instead of a polarization of pure horizontal character (dashed dark blue). Right panel: Corresponding voltage loops of the oscillator strength, S, of the field-induced TO mode. (d) Temperature dependence of the $LO_4$ anomaly. Left panel: Spectra of the field-induced TO mode (cyan) together with the fit (dashed purple) at selected temperatures. Right panel: Temperature dependence of S for the field-induced TO mode (at -8kV/cm).*

Complementary information about the interfacial polar moments has been obtained from a related phonon anomaly at the longitudinal optical ($LO_4$) edge of STO around 800 cm$^{-1}$ that is shown in Fig. 1c. It arises from a transverse optical (TO) mode that is induced by the polar distortion and is infrared active only in the direction parallel to $P$ (details are given in section 5 and 6 of the SI).

Near such an LO edge, the ellipsometric response is sensitive to both the horizontal and the vertical components of the dielectric function [36], [37], [38] (for details see section 3 and 7 of the SI). Accordingly, the analysis of this $LO_4$ anomaly yields further information about the vertical component of $P$ (in addition to the one from the R-mode splitting on the horizontal component). Figure 1c reveals that the $LO_4$ edge anomaly cannot be reproduced if $P$ is predominantly laterally orientated with $P_{ab} \gg P_c$ (dashed dark blue line), whereas it is well described assuming that $P$ is oriented along a diagonal with respect to the surface normal with $P_{ab} = P_c$ (dashed purple line).

The right panel of Fig. 1c displays the gate-field loop of the amplitude of the transverse mode at the $LO_4$ edge, as obtained with the diagonal polarization model ($\boldsymbol{P}_{ab}=\boldsymbol{P}_c$), which shows a similarly asymmetric behavior as the R-mode splitting in the top-right panel of Fig. 1a. This confirms the common origin of these phonon anomalies and suggests that the underlying polar moments are strongly (but not fully) inclined toward the interface. A likely scenario is that $\boldsymbol{P}$ is directed along a diagonal with respect to the Ti-O bonds (or the surface normal), similar as in orthorhombic or rhombohedral bulk $BaTiO_3$ [39].

Finally, Fig. 1d displays the temperature dependence of the $LO_4$ edge anomaly which shows that the interfacial polar moments persist to much higher temperature than the field-induced ferroelectric order in bulk STO [12] and vanishes along with the AFD domain boundaries near $T^*=105$ K.

**Oxygen vacancy clustering scenario**

A scenario which can account for the above described asymmetric and non-collinear interfacial polar moment, $\boldsymbol{P}$, and its close link with the AFD domain boundaries of STO is sketched in Fig. 2. Here, the positively charged oxygen vacancies are assumed to migrate in response to the gate voltage, whereby they get trapped and cluster at the AFD domain boundaries. The latter act as extended pinning centers that are strengthened (weakened) at -V (+V).
In the pristine state (before the gate voltage is applied), the oxygen vacancies are more or less randomly distributed in the vicinity of the interface and probably weakly pinned at some local defects (see Fig. 2b). A positive gate voltage (+V) pushes the positive oxygen vacancies toward the interface (see Fig. 2c). It also increases the concentration and thickness of the 2DEG layer, which thus screens most of the oxygen vacancies and prevents the formation of extended clusters and subsequent polar moments close to the interface. Accordingly, as shown in Fig. 2c, the induced polarization at large +V has a predominant vertical orientation and develops only at a rather large distance from the interface that exceeds the probing depth of the infrared ellipsometry experiment (of about one micrometer, as shown in section 9 of the SI). Note that whereas the majority of the confined electrons reside within a few nanometers from the interface [40], their depth profile has a long tail that can extend over hundreds of nanometers (especially at +V) [41], [42] due to the large dielectric constant of STO.

As the gate voltage is subsequently reduced toward -V, the 2DEG becomes depleted and its thickness decreases. Hence, the positive oxygen vacancies, as they start moving away from the interface, get trapped at the AFD domain boundaries which thus become charged and give rise to local electric fields and induced polar moments with sizeable lateral components (see Fig. 2d) that are close to the interface and thus readily seen in the infrared ellipsometry spectra. Note that an additional contribution to these local fields may arise from the flexoelectric behaviour of STO that is strongly enhanced by the large low-T dielectric constant [19]. As shown in Fig. 2e-g, the clustering of the oxygen vacancies causes an increase of the lattice parameter [43] and a subsequent strain gradient across the AFD domain boundaries. The latter creates a flexoelectric field, $\boldsymbol{E}^{flexo}$, which is parallel to the Coulomb-field, $\boldsymbol{E}^{Coul}$, and therefore helps to induce interfacial polar moments that are strongly asymmetric and non-collinear with respect to the applied gate field.

Note that at large -V the 2DEG may also become laterally inhomogeneous [44] and thus may further enhance the above described local electric fields (see section 12 of the SI).

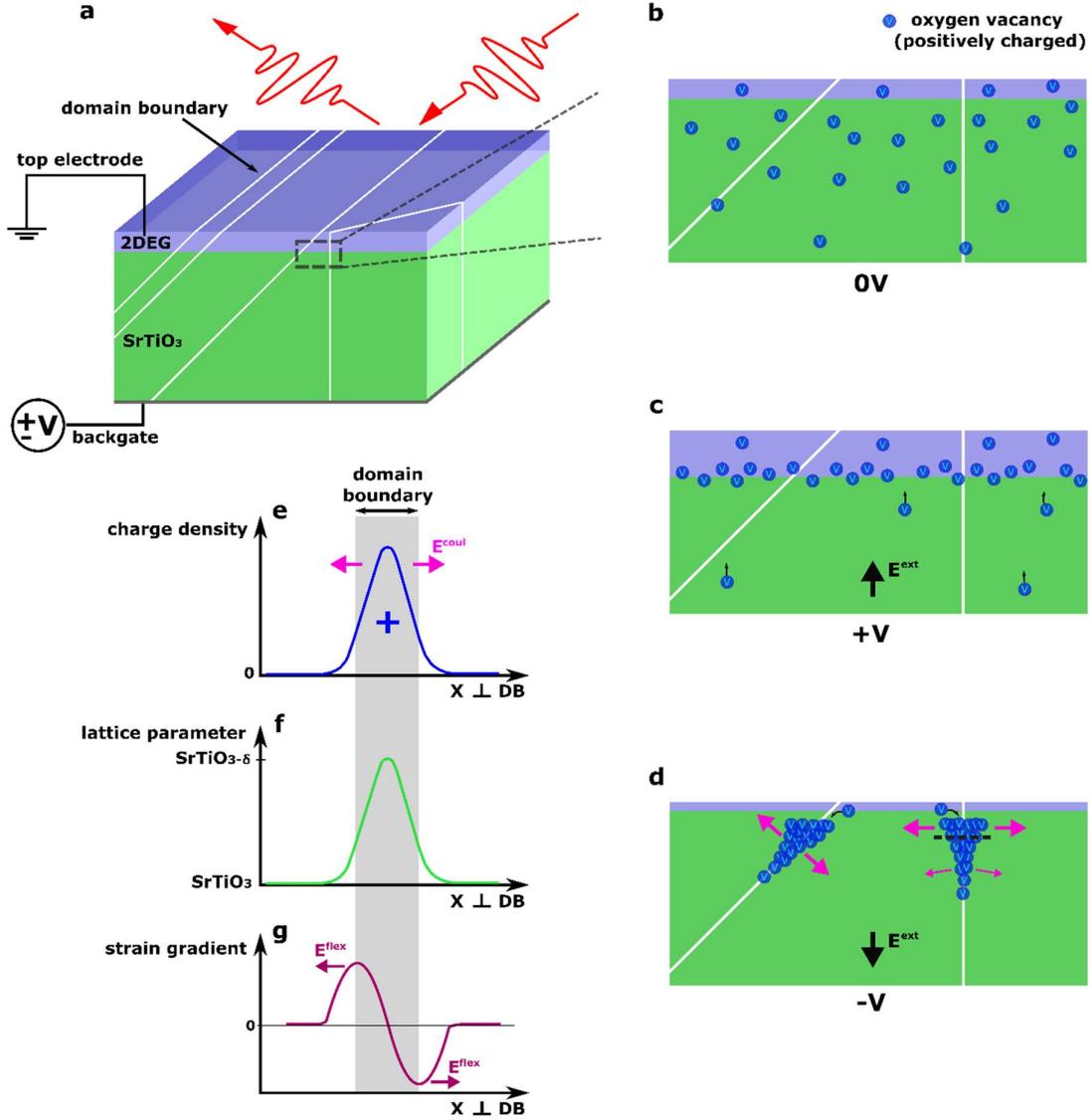

*Figure 2: Sketch of a field-gated STO-based heterostructure, showing the electromigration of oxygen vacancies and their clustering at the AFD domain boundaries, which gives rise to the anomalous horizontal polarization, $P_{ab}$, at -V. (a) Sketch of the heterostructure and its gating between the 2DEG and a silver contact on the backside of STO. The white lines show the AFD domain boundaries that appear at T < T\*. (b) Magnified view of the 2DEG (purple) and the distribution of the positively charged oxygen vacancies (blue) in the vicinity of the interface in the pristine state at 0V. (c) A positive gate voltage +V pushes the positive oxygen vacancies towards the interface and thereby induces a polarization and an electric field (magenta arrows) that is perpendicular to the interface, i.e. $P_c > 0$. Note that the additional contribution due to the electric-field-induced ferroelectric polarization of bulk STO is not shown (this is also valid for (d)). (d) At -V the oxygen vacancies are pulled away from the interface and get preferably trapped at the AFD domain boundaries. The resulting electric charging and the flexoelectric effect induced by the strain gradients due the oxygen clustering create a polarization with a large horizontal component, $P_{ab}$, which is strongest close to the interface. (e,f,g) Cross-section of a domain boundary at -V (black dashed line in d). The accumulated, positively charged, oxygen vacancies create an electric field $E^{coul}$ (e) as well as and an expansion of the lattice parameter of $SrTiO_3$ (f). This induces a strain gradient (g) and thus strong flexoelectric fields $E^{flex}$, which are perpendicular to the domain boundaries.*

As shown in Fig. 2d, the direction of the induced polar moments depends on the type of domain boundaries, i.e. $\boldsymbol{P}$ is mainly laterally oriented for the boundaries between the x- and y-domains (denoted as x/y-boundaries) whereas it is at 45 degree with respect to the surface normal ($\boldsymbol{P}_{ab} \approx \boldsymbol{P}_c$) for the x/z- and y/z-boundaries [45]. The above described results therefore suggest a preferred clustering of the oxygen vacancies at the x/z- and y/z-boundaries.

**Raman spectroscopy**

Complementary information on the gate-voltage-induced polar moments has been obtained with confocal Raman spectroscopy, which has a larger probing depth than infrared ellipsometry and enables scans of the polarization depth profile [46], [47]. Figure 3a shows a sketch of the Raman experiment in grazing incidence geometry with the sample mounted on a wedged holder. The red laser beam ($\lambda = 633$ nm) is incident at 70° to the surface normal and the refracted beam inside STO with $n \approx 2$ is near 30° [48], [49]. The electric field vector $\boldsymbol{\epsilon}$ of the light has therefore a sizeable (zero) vertical component in p-polarization (s-polarization).

Figure 3b shows the Raman spectra of AlO$_x$/STO at 10 K with incoming p-polarization at 0 V and ±4 kV/cm. The background with several broad maxima, arising from multi-phonon excitations, is characteristic for STO (e.g. [5], [50], [51]) and hardly affected by the gate voltage. At 0 V (black line) the direct phonon excitations are due to R-modes at 15, 45, 144, 229 and 447 cm$^{-1}$ (black stars) that become Raman-active below $T^* = 105$ K (the cubic phase has no Raman-active phonons) [5], [50], [4]. At ±4 kV/cm, several additional peaks develop around 25, 175, 540 and 795 cm$^{-1}$ (green arrows) which correspond to the infrared-active TO$_1$, TO$_2$, TO$_4$ and LO$_4$ modes (the TO$_3$ peak at ~263 cm$^{-1}$ is barely resolved) that are activated by a polar distortion which breaks the inversion symmetry [51].

Figure 3c magnifies the low energy range with the TO$_1$ soft mode that exhibits the largest field-effect. The inset shows corresponding s-polarized spectra (normalized to the multi-phonon background) for which the soft mode intensity is significantly weaker. Considering that the Raman intensity of the soft mode is expected to be maximal for $\boldsymbol{\epsilon} \parallel \boldsymbol{P}$ and minimal (but finite) for $\boldsymbol{\epsilon} \perp \boldsymbol{P}$, the polarization dependence in Fig. 3c is therefore consistent with the above described scenario of induced polar moments that are oriented along the diagonal (vertical) direction at -4 kV/cm (+4 kV/cm). Note that a quantitative analysis with respect to the orientation of $\boldsymbol{P}$ would require a much more extensive polarization study and a lateral scanning, as to resolve individual domains with different orientation of $\boldsymbol{P}$, which is beyond the scope of this study.

Next, we discuss the depth dependence of the soft mode intensity at ±4 kV/cm in Figure 3d. At +4 kV/cm, the soft mode intensity shows only a weak variation that agrees with the scenario of a vertical and laterally homogeneous polarization that extends deep into the STO substrate (see Fig. 2c). Close to the interface, where the ellipsometry spectra indicate the absence of polar moments since they are screened by the enhanced 2DEG layer, the soft mode intensity exhibits indeed a clear decrease. The circumstance that this decrease is only partial can be understood in terms of the probe depth of the confocal Raman experiment, which is of the order of several micrometers and therefore larger than for infrared spectroscopy.

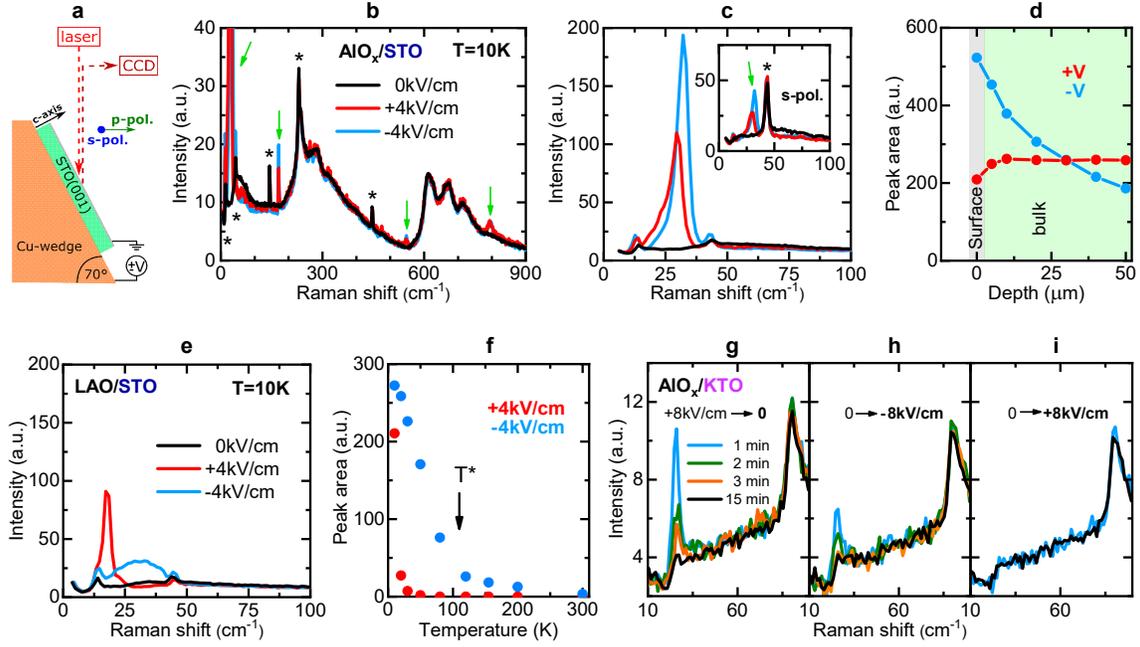

*Figure 3: Confocal Raman study of the induced polar moment, as a function of gate voltage, distance from the interface, and temperature. (a) Sketch of the measurement geometry. (b) Gate-voltage dependent spectra of an AlO$_x$/SrTiO$_3$ heterostructure at 10 K. A Stokes/Anti-Stokes comparison yields a temperature offset of max. ~20 K, caused by laser heating. Black stars mark the R-modes at 15, 45, 144, 229 and 447 cm$^{-1}$ that become Raman-active below T\* = 105 K. Green arrows show the field-induced infrared-active phonon modes at 25, 175, 540 and 795 cm$^{-1}$. (c) Magnified view of the field-induced soft mode (TO$_1$) and its asymmetry with respect to ±V. Inset: Corresponding spectra taking in s-polarization (normalized to the multi-phonon background). (d) Depth profile of the normalized peak area of the soft mode for +V and -V. (e) Field-dependence of the soft mode of an LAO/STO structure. (f) Temperature dependence of the peak area of the soft mode for +V and -V. (g-i) Low-frequency Raman spectra of an AlO$_x$/KTaO$_3$ heterostructure, showing a weak and rapidly decaying (within minutes) soft mode signal that occurs only after the voltage has been decreased, i.e. either from +V to 0V (g) or from 0V to -V (h), but is absent after a corresponding voltage increase from 0V to +V (i) or from -V to 0V (not shown).*

The corresponding depth scan at -4 kV/cm highlights a remarkably different trend. Here, the soft mode intensity is strongly enhanced at the interface but decreases rapidly toward the bulk of STO, where it becomes comparable to the one at +V. Similar soft mode intensities at ±V in the bulk of STO are also seen in a conventional macro-Raman experiment that probes deep into the STO substrate (see section 13 of the SI). These characteristic differences in the depth profile of the gate-voltage-induced polarization agree with the scenario sketched in Fig. 2d. At –V, the polar moments close to the interface are strongly enhanced by the local fields due to the charged and flexoelectric AFD domain boundaries and, in addition, are barely screened by the depleted 2DEG.

Figures 3c and 3e show a comparison of the gate-voltage-induced Raman soft modes of the AlO$_x$/STO and LAO/STO heterostructures. Whereas the field-induced soft mode peaks are similar for both samples at +4 kV/cm (red curves), they are dramatically different at -4 kV/cm (cyan curves) where the peak is much stronger and sharper in AlO$_x$/STO than in LAO/STO. This trend can be understood in terms of the higher oxygen vacancy concentration in AlO$_x$/STO, which gives rise to larger clusters at the AFD domain boundaries and, correspondingly, to larger local electric fields. The mobility of the oxygen vacancies might also depend on the strain that is imposed by the top layer. For the amorphous AlO$_x$ layer, the strain is weakly tensile and thus more favorable for creating large oxygen vacancy clusters than the strongly compressive strain of the epitaxial LAO layer.

Another difference between the induced polarizations at ±4 kV/cm is evident in Fig. 3f from the temperature dependence of the Raman soft mode intensity (here shown for LAO/STO). At +4 kV/cm, the soft mode intensity decreases rapidly and vanishes above 40 K, in accordance to the electric-field-induced ferroelectric order in bulk STO [12]. The soft mode at -4 kV/cm, however, persists to about $T^*=105$ K where it vanishes together with the AFD domain boundaries. The weak signal above $T^*$ arises most likely from remnant AFD surface domains that persist well above $T^*$ [52].

Figures 3g-i show a corresponding Raman study of an $AlO_x/KTaO_3$ (001) heterostructure which corroborates the central role of the AFD domain boundaries in the above described interfacial polarization phenomena of the STO-based structures. KTO is also a quantum paraelectric with a diverging $\varepsilon_0$ at low temperature, albeit not as close to the ferroelectric critical point as STO [53], [54]. Unlike STO, the structure of KTO remains cubic down to the lowest temperature such that in the quantum paraelectric regime it has no AFD domains. The KTO-based devices also host a 2DEG [55], [56], [57], [58] that can be readily modified with a back-gate voltage, although with less pronounced carrier localization effects at -V than in the STO devices [59]. Figures 3g-i show that in the Raman spectra of $AlO_x/KTO$ the back-gate voltage induces only a weak and short-lived soft mode peak. The metastable soft mode vanishes within minutes and has a similar intensity in s- and p-polarization (see section 14 of the SI). Importantly, the induced peak appears only after the gate voltage has been decreased, i.e. after reducing the field from +8 kV/cm to 0 V or from 0 V to -8 kV/cm, but not after an increase from -8 kV/cm to 0 V or from 0 V to +8 kV/cm. This behavior is consistent with the oxygen vacancies becoming weakly pinned by some local defects, which, unlike the AFD domain boundaries of STO, are not extended and thus not oriented along a particular direction. In return, this highlights the central role of the AFD domain boundaries for the strongly asymmetric and non-collinear polarization behavior of the STO-based devices.

**UV illumination**

Figure 4 demonstrates that the anomalous interfacial polarization of an $AlO_x/STO$ device can be readily modified with other external stimuli, such as UV light. Figure 4a reveals that the splitting of the R-mode in the ellipsometry spectra at -8 kV/cm (cyan line) can be fully suppressed upon UV illumination (violet line) and remains almost completely absent even after the UV light has been switched off (orange line). The original R-mode splitting can be restored by cycling the gate voltage to +8 kV/cm (red line) and then back to -8 kV/cm (dark blue line). Fig. 4c confirms that the same kind of optical 'switch-off' effect occurs for the gate-field-induced anomaly at the $LO_4$ edge. The underlying mechanism indicate that photo-generated charge carriers (across the band gap of STO) enhance the population of the 2DEG and thus make the screening of the oxygen vacancy clusters more efficient. Figures 4b and 4d demonstrate that the UV light can also erase the remnant polarization (and thus the memory) at 0 V after a complete electric field cycle.

We expect that this demonstration of an electric 'switch on' and optical 'switch off' mechanism of the interfacial polarization is only one of potentially many examples of how the local and extended defects of STO and their interplay with the 2DEG can be used for engineering STO-based devices with improved or new functionalities (see e.g. Ref. [60]). Further promising candidates are external pressure, internal strain, chemical substitution, doping and different surface cuts of STO.

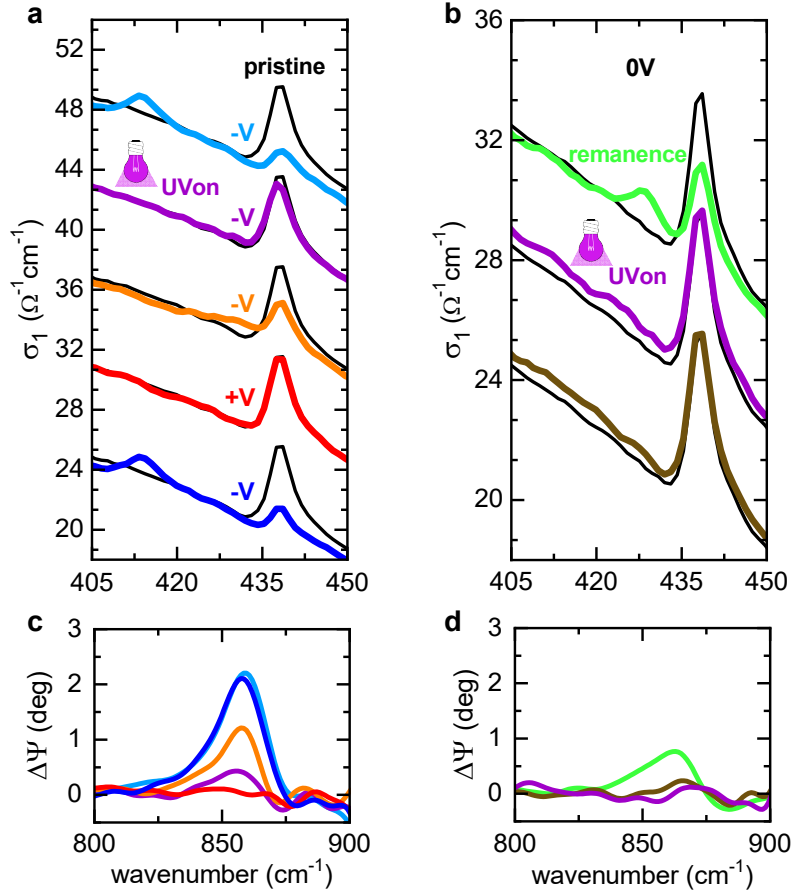

*Figure 4: Optical switching of the anomalous polarization in $AlO_x/SrTiO_3$ at 10 K. (a) Infrared spectra showing that the pronounced R-mode spitting at -8 kV/cm (cyan) can be erased by UV illumination (violet) and is only partially restored after the illumination (orange). The original R-mode splitting is restored after a field cycle to +8 kV/cm (red) and back to -8 kV/cm (blue). (b) Infrared spectra at 0V showing that the remnant R-mode splitting (green) is erased by the UV light (violet) even after it has been turned off again (brown). (c,d) Corresponding infrared spectra of the polarization induced TO mode at the $LO_4$ edge. Shown are the difference spectra with respect to the pristine state at 0V, i.e. $\Delta\Psi=\Psi_{exp} – \Psi_{0V, pristine}$. The data are smoothened with a FFT filter.*

**Conclusions**

In summary, with infrared ellipsometry and confocal Raman spectroscopy we have shown that electric back-gating of LAO/STO and AlO$_x$/STO heterostructures gives rise to interfacial polar moments that are non-collinear and strongly asymmetric with respect to the vertical gate field. Our results provide evidence for an important role of oxygen vacancies which tend to form extended clusters at the AFD domain boundaries that give rise to electric charging and flexoelectric effects. The subsequent interfacial polar moments with large horizontal components are induced at lower gate voltages and persists to much higher temperatures than in bulk STO.

The training and hysteresis effects due to these anomalous interfacial polar moments are reflected in the magneto-transport properties of related devices [29], [28]. However, since the laterally inhomogeneous polarization effects tend to be cancelled out in vertical structures, their real potential can only be exploited with lateral device structures. The control of the AFD domain boundaries and their interaction with the oxygen vacancies likely enables new device concepts and functionalities. Recently, there have indeed been great advances in patterning ferroelectric domains [61], which may also be effective in STO due to its polar nature.

**Acknowledgments**

F.L. and C.B. acknowledge enlightening discussions with S. Das, J. Maier, R. Merkle, A. Dubroka, and B. I. Shklovskii. Work at the University of Fribourg was supported by the Schweizerische Nationalfonds (SNF) by Grant No. 200020-172611. M.B. acknowledges support from the ERC Advanced grant n° 833973 "FRESCO" and the QUANTERA project "QUANTOX". GH acknowledges financial support from Spanish Ministry of Science, Innovation and Universities, through the "Severo Ochoa" Programme (CEX2019-000917-S) and the MAT2017-85232-R (AEI/FEDER, EU) projects, and Generalitat de Catalunya (2017 SGR 1377).

**Author contributions**

F.L. performed and analyzed the infrared ellipsometry measurements under the direction of P.M., M.Y.-R., B.X., Y.P. and C.B.. The confocal Raman spectroscopy measurements were performed and analyzed by F.L. with the help from M.M., K.F., B.K., B.P.P.M, A.C., A.S., Y.P. and C.B.. The macro-Raman measurements were carried out by M.C. and P.H.. L.M.V.A., D.C.V and M.B. grew the samples and performed the magneto-transport measurements. C.B. and F.L. conceived and planned the project, and wrote the manuscript with inputs from all authors. All authors discussed the results.